\documentclass[%
 reprint,
 amsmath,amssymb,
 aps,
]{revtex4-2}

\usepackage{graphicx}
\usepackage{dcolumn}
\usepackage{bm}

\usepackage{xcolor}

\newcommand{\pcc}{ cm$^{-3}$ }

\newcommand{\ixpe}{\textit{IXPE} \,}
\newcommand{\IXPE}{\textit{IXPE} \,}
\newcommand{\chandra}{\textit{Chandra} \,}

\newcommand{\kms}{km~s$^{-1}$}

\def\lsim{\;\raise0.3ex\hbox{$<$\kern-0.75em\raise-1.1ex\hbox{$\sim$}}\;}
\def\gsim{\;\raise0.3ex\hbox{$>$\kern-0.75em\raise-1.1ex\hbox{$\sim$}}\;}
\def\apj{ApJ}
\def\mnras{MNRAS}

\def\aap{A\&A}                   
\def\aj{AJ}                      
\def\apjs{ApJS}                  
\def\apjl{ApJ}                   

\def\ssr{Space Sci. Rev.}
\def\aapr{Astron. Astroph. Reviews}
\def\physrep{Phys. Reports}

\def\apj{ApJ}
\def\apjl{ApJ Lett.}
\def\mnras{MNRAS}

 \def\aap{Astron. Astrophys.}                  
\def\aj{Astron. J.}                      
\def\apjs{Astrophys. J. Suppl. Ser.}                  
\def\apjl{ApJ Lett.}                   

\def\ssr{Space Sci. Rev.}

\def\aapr{ Astron. Astrophys. Rev.}

\usepackage{url}
\usepackage{xcolor}
\usepackage{ulem}
\definecolor{newcolor}{rgb}{.8,.349,.1}

\begin{document}
\preprint{APS/123-QED}

\title{X-ray polarization: A view deep inside cosmic ray driven turbulence and particle acceleration in supernova remnants}

\author{Andrei.M. Bykov}
\email{byk@astro.ioffe.ru}
\affiliation{Ioffe Institute, 194021, Saint-Petersburg, Russia}
\author{Donald.C. Ellison}
\email{ellison@ncsu.edu}
\affiliation{Physics Department, North Carolina State University, Box 8202, Raleigh, NC 27695, USA}
\author{Sergei.M. Osipov}
\email{osm.astro@mail.ioffe.ru}
\affiliation{Ioffe Institute, 194021, Saint-Petersburg, Russia}
\author{Patrick Slane}
\email{slane@cfa.harvard.edu}
\affiliation{Harvard-Smithsonian Center for Astrophysics, Cambridge MA 02138, USA}
\author{Yury.A. Uvarov}
\email{uv@astro.ioffe.ru}
\affiliation{Ioffe Institute, 194021, Saint-Petersburg, Russia}

\date{\today}

\begin{abstract}
We show here that highly polarized X-ray synchrotron radiation from young supernova remnants (SNRs) can be modeled within the framework of diffusive shock acceleration (DSA) and nonlinear magnetic turbulence generation. Cosmic ray acceleration by SNR 
shocks to very high energies requires efficient magnetic turbulence amplification in the shock precursor. 
As the strong turbulence generated by Bell's instability
far upstream from the viscous subshock convects through the subshock, nonlinear dynamical effects on the 
large amplitude, compressible fluctuations produce a downstream layer filled with strong anisotropic turbulence 
with predominantly radial magnetic fields.
The synchrotron radiation from shock accelerated electrons in the turbulent downstream layer has a 
high degree of polarization shown to be consistent with recent observations of young SNRs by 
the {\sl Imaging X-ray Polarimetry Explorer (IXPE)} taking into account high-energy electron losses and line-of-sight integration in a spherical remnant.
In the case of our model of Tycho's SNR, 
the measured X-ray radiation constrains  
the thickness of the energy containing interval and 
the amplitude of cosmic ray driven magnetic turbulence, as well as the maximal energy of accelerated protons. The preferential direction of the X-ray polarization depends sensitively on the SNR shock velocity and the ambient density. A fast shock in a region with high enough density is a favorable place to produce tangential polarization of synchrotron radiation, i.e., a dominantly radial turbulent magnetic field.  
A unique feature of our model is the sensitive dependence of the degree and direction of X-ray polarization
on the spatial overlap between regions of 
amplified magnetic turbulence and TeV electron populations. While this overlap occurs on scales orders of magnitude  below
the resolution of {\sl IXPE}, its polarization measurement allows testing of turbulent plasma processes on 
unprecedented scales.The mechanism of formation of highly polarized X-ray synchrotron radiation in fast shocks with high level of anisotropic turbulent magnetic field preferentially directed along the shock normal may be applied to other systems like 
shocks produced by black hole jets. 
\end{abstract}

\maketitle


\section{Introduction}
Radio, X-ray, and gamma-ray observations have proved undoubtedly that young supernova remnants (SNRs) are  cosmic ray (CR) accelerators \citep[][]{helder12, 2013A&ARv..21...70B, 2014IJMPD..2330013A,2014NuPhS.256....9B,2018SSRv..214...41B}, while there are still important unresolved questions concerning details of the physical processes involved.
Accelerated relativistic electrons  emit synchrotron radiation in SNR magnetic fields \citep{Ginzburg1965} which is detected in radio and X-rays in young SNRs. 
High angular resolution \chandra observations provided clear evidence for strong, 
non-adiabatic amplification of magnetic fields in the vicinity of 
SNR forward shocks \citep[e.g.][]{Parizot2006,helder12}. Detection and mapping of polarized synchrotron  radiation is a powerful observational tool to probe the structure of magnetic fields. This can be performed in the X-ray \citep[see e.g][]{Bykov_Bloemen2009,Baring2017a,2020ApJ...899..142B} and radio bands \citep[see e.g.][]{2016MNRAS.459..178B,2017MNRAS.470.1156P}. 
Radio polarization observations of Tycho \citep{1991AJ....101.2151D,1997ApJ...491..816R} 
and a few other young SNRs \citep{1987AuJPh..40..771M,1993AJ....106..272R}, revealed magnetic fields that are predominantly radial at the SNR blast wave rim.
The degree of polarization in the limb brightened rims of Tycho reach 20\%-30\%, while it is about 7\% in the main shell observed at a wavelength of 6 cm \citep{1991AJ....101.2151D}. 
The radio polarization in Tycho's SNR shows a relatively high degree of polarization (DP)
\citep{1992AJ....103.1338W} for which the magnetic field is organized in loose cell-like structures with a maximal scale size of about 110 arcseconds (1.3 pc).

The modern {\it Imaging X-ray Polarimetry Explorer} ({\sl IXPE}) spacecraft, using photoelectric detection technologies \citep{ixpe_soffitta21,ixpe_w22}, recently discovered polarized X-ray radiation from 
the young SNRs Cas A \citep{ixpe_CasA22}, Tycho \citep{ixpe_Tycho23}, and SN1006 \citep{ixpe_SN1006}. 
With the \ixpe spatial resolution ($\sim30$ arcseconds) it was possible to measure the DP from a few localized sub-parsec scale regions in the Western part of Tycho with the maximal DP of 23\% $\pm$ 4\%. 
The DP averaged over the whole remnant is 9\% $\pm$ 2\% . Significantly, the direction of the observed  polarization is mainly tangential to the shock, a signature of  preferentially 
radial magnetic fields in the X-ray synchrotron emitting region. 

The observation of polarized X-ray synchrotron radiation in SNRs indicates the presence of electrons accelerated above TeV energies in amplified, turbulent magnetic fields. The most likely explanation for this electron acceleration is nonlinear diffusive shock acceleration (DSA) driving Bell's instability for the generation of strong magnetic turbulence \citep[see e.g.][]{BE87,Malkov_Drury_2001,Bell04,helder12}.
At first glance, the high degree of  X-ray polarization measured by \ixpe is difficult to understand since the strong magnetic turbulence required by efficient DSA should limit the polarization.  
Here, we describe a model that can consistently account for the observed polarization and efficient particle acceleration with nonlinear DSA.
We emphasize that the observation of 
such high polarization is an important clue to the nature of structured turbulent magnetic fields deep inside the remnant, as well as the very high energy particle distribution.

A number of magnetic field amplification mechanisms have been 
considered for SNRs including, in particular, 
the Rayleigh Taylor instability \citep[see e.g.,][for a MHD model]{1996ApJ...465..800J,BE2001}.
The observation of polarised X-ray emission adds an important constraint on any amplification model by adding information on the morphology of the magnetic field as well as the strength. 
Here we present a model of polarized X-ray synchrotron emission from Tycho's SNR where 
CR driven instabilities produce the 
magnetic turbulence amplification \citep{BE87,Malkov_Drury_2001,Bell04,Zirakashvili_2008pol,Amato09,SchureEtal2012,Bykov3inst2014}.

The bright narrow rims of non-thermal X-ray emission revealed in  \chandra images in the vicinity of the forward shocks of a few young SNRs is clear evidence for 
amplification of magnetic fields by a factor of 50-60 or more, well above that expected from MHD processes. 
Diffusive shock acceleration is the most promising mechanism of 
CR acceleration in SNRs \citep[e.g.,][]{LC83,BE87,Achterberg94}. 
The strong nonlinear nature of DSA, coupled with the generation of large CR currents in the shock precursor, naturally supports Bell's instability for producing magnetic turbulence strong enough to allow SNR shocks to accelerate CRs to well above TeV energies. At lower 
energies the stochastic re-acceleration of radio-emitting electrons in the shock downstream may play a role \citep{2020A&A...639A.124W}.

\section{Model description}
In this paper we model the spatial structure of anisotropic magnetic turbulence together 
with the acceleration of multi-TeV electrons producing synchrotron photons.
Our model assumes 
that thin X-ray filaments observed in young SNRs are 
synchrotron radiation produced by multi-TeV electrons accelerated by DSA at the remnant forward shock.
Models of DSA in young SNRs require a high level of magnetic turbulence of rather short scales
in the shock precursor (gyro-scales of CRs are typically well below 0.1 pc).
It has been shown that 
CR proton driven instabilities \citep[e.g.,][]{BE87,Malkov_Drury_2001,SchureEtal2012}
can amplify seed circumstellar magnetic fluctuations to provide this turbulence.
Specifically, Bell's non-resonant instability, driven by CR currents in the shock precursor, 
was shown to provide the required high level of turbulence amplification \citep{bell_lucek01,Bell04,Zirakashvili_2008pol,Bykov3inst2014,Caprioli_MFA14}.
The turbulence has a slightly smaller longitudinal component of the magnetic field than transverse.
Moreover, the non-linear phase of Bell's turbulence produces 
strong plasma density fluctuations on CR gyro-scales. 

The upstream plasma density fluctuations convect into the viscous subshock and produce 
shock surface ripples.
These ripples in turn efficiently produce  turbulence behind the shock  \citep{1987flme.book.....L,Bykov1982,GiacJok07,2013ApJ...770...84F,Lemoine_corrugation16,Trotta_BStransm22}.

The downstream turbulence is anisotropic where the RMS direction 
of the fluctuations is preferentially longitudinal, i.e., perpendicular to the shock surface \citep{Zirakashvili_2008pol,inoue13,lazarian_MFA22,federath23}.

The physical mechanism we consider for amplification of the downstream magnetic field  has three main phases. 
First, CR protons accelerated by a strong collisionless shock stream into the upstream shock precursor where their current 
amplifies seed magnetic turbulence by nonlinear CR-driven instabilities \citep[e.g.][]{Bell04,SchureEtal2012}. 
We specifically consider the Bell instability generated by the high-energy CR proton current leaving the shock far upstream.
The same DSA process accelerates ambient electrons in young SNRs to above TeV energies. 
Second, the plasma density perturbations produced during the non-linear evolution of Bell's magnetic turbulence in the CR precursor convect into the viscous subshock and produce
ripples in the subshock surface. 
These ripples (corrugations) efficiently produce intense plasma fluctuations, containing 
a sizeable fraction of the turbulent energy, that propagate 
downstream from the subshock \citep[see e.g.][]{1987flme.book.....L,Bykov1982,1986MNRAS.218..551A,2007ApJ...663L..41G}.
The shock corrugations may also affect the energetic particle injection process 
\citep[see e.g.][]{2021FrASS...8...27G}.

Third, as the  turbulence convects, it produces intense, 
anisotropic magnetic turbulence in a layer just behind the shock. The amplification process is the small scale turbulent dynamo mechanism 
described in \citep{1968JETP...26.1031K,2005PhR...417....1B,SHUKUROV2008251,2016ApJ...833..215X,Shukurov_Subramanian_2021}.
Importantly, the character of the magnetic field anisotropy changes with distance downstream.
The transverse component (i.e. along the shock surface) of the magnetic field is 
amplified by the shock compression and it dominates over the longitudinal one in the turbulent field immediately after the shock jump. 
However, the anisotropic vortex turbulence  (dominating the total energy density of turbulence in the downstream)  preferentially amplifies the longitudinal magnetic field component by the fast small-scale dynamo mechanism.

Thus, at a distance from the shock on the order of the characteristic scale of the upstream turbulence, 
the downstream RMS turbulent magnetic field may switch polarities and become dominated by the longitudinal field component resulting in transverse
polarized synchrotron radiation from relativistic electrons.
We defined here the characteristic scale of the CR driven upstream turbulence at the peak of the energy containing interval in Fig. \ref{energy_sp_Bell}.
Farther downstream, the dynamo produced magnetic turbulence decays in the absence of driving sources, in accordance with MHD models. The RMS amplitudes of the longitudinal (green line, $B_r$)  and the transverse (red line, $B_\mathrm{tr}$) components of the turbulent magnetic field are shown in Fig.~\ref{fig:sketch}.

The strong amplification of the turbulent magnetic field in the shock vicinity increases the 
synchrotron losses of relativistic electrons. The highest energy electrons, 
which radiate the X-ray synchrotron photons in young SNRs, are distributed in a narrow layer 
around the subshock (see for a review \citep{helder12}), as shown in Fig.~\ref{fig:sketch}  by the violet lines.
Lower energy radio-emitting electrons would have a broader distribution. 
The observed direction of X-ray synchrotron polarization depends on the width of the electron distribution 
together with the amplitude of the turbulent field (green and red curves in Fig.~\ref{fig:sketch}).
If the high energy electron distribution is wide enough to 
extend farther behind the shock than the peak of the green line (bottom panel in Fig.~\ref{fig:sketch}), 
the polarization direction will be transverse, 
as is indeed observed in X-ray observations of Tycho's SNR, 
and in many radio images of young SNR as mentioned above.   
If the case with a narrow electron distribution occurs 
(top panel of Fig.~\ref{fig:sketch}),
the X-ray polarization will be longitudinal.

To demonstrate our model, 
we choose parameters typical of Tycho's SNR. 
Tycho is an extended young remnant for which deep X-ray ({\sl Chandra}) and polarimetric ({\sl IXPE}) 
observations are available, as well as extensive multi-wavelength data. 
An analysis of Tycho's expansion rate \citep{williams16} derived the maximum forward shock velocity  of $\sim$ 5,300 \kms, assuming the 
distance to Tycho to be 2.3-4 kpc. 

Indications of order-of-magnitude variations in the ambient density around the periphery of Tycho's forward shock, possibly due to interactions with dense clumps of the interstellar medium, were found by \citep{williams13}. 
While the mean ambient number density  is about \hbox{0.1-0.2 \pcc,} they found that in the Western regions it may be 3-10  compared to that in the southwest (see also \citep{2023ApJ...951..103E,2024ApJ...961...32K} for recent discussions). 
The map of polarized X-ray emission of Tycho's SNR by \citep{ixpe_Tycho23} revealed a high degree of polarization in the northeast region. 
To model  synchrotron X-ray radiation from different regions we consider a range of ambient number densities below 1 \pcc.

\section{Intensity and polarization simulation}

The intensity and polarization of Tycho's synchrotron map can be obtained
after integration of the Stokes emission parameters over the line-of-sight (LOS)
using formulae for synchrotron radiation obtained in 
\citep{Ginzburg1965}. 
The details are in Appendix A.
Briefly: as the plasma flows downstream from the subshock, 
the evolution of the turbulent magnetic field and the electron distribution function are calculated. The strong synchrotron losses produce a spatial  inhomogeneity where the radiation is concentrated in a thin layer just downstream from the subshock. 
However, general formulae have been simplified using some assumptions of the problem symmetry and averaging. 

The magnetic field simulation is discussed in Section \ref{sec:setup}. The field was calculated for a plain shock geometry
in a cubic box of size $\sim10^{17}$ cm (see Section \ref{sec:setup}). 
Since the Tycho SNR radius is $\sim10^{19}$ cm, larger than our box size, 
we must extrapolate and adjust for the spherical remnant to calculate the polarization
degree and emission radial profile across the remnant.

First of all, we construct smooth analytical approximations for square-averaged magnetic fields
neglecting small scale fluctuations, as described in Appendix \ref{sec:fitting}. 
The approximation has a different
functional dependence on distance from the shock for radial ($B_{r}$) and transverse (${B_\mathrm{tr}}$) field components.
$B_{r}$ is directed along the SNR radius while $B_\mathrm{tr}$ is normal to $B_{r}$. We assume spherical symmetry and local axial
symmetry of an anisotropic magnetic field where the direction of $B_{r}$ is the axis of symmetry. 
While integrating along the LOS we average  over the
transverse magnetic field direction (and over small scale
magnetic fluctuations, see Appendix A) at each local point. 

Diffusive shock acceleration  produces an approximate power law distribution
of high energy electrons in a broad energy band with an
exponential cut off at some energy. 
However, the synchrotron X-ray photons observed in the \hbox{$1-8$~keV} range were emitted by electrons in a  
narrow energy interval  above \hbox{1 TeV.} 
Therefore, even in the cut-off regime,  one can approximate the electron distribution function 
for the observed radiation as a power law $f_{e}\left(E\right)=K_{e}E^{-p}$.
In this case the integration over energy can be done analytically and the local averaging over stochastic magnetic fields described in Appendix A can be done after it. After integration along the LOS a 1-dimensional
SNR intensity and polarization degree radial profile can be obtained.

The real distribution of the fluctuating magnetic field in the synchrotron emitting layer may differ from the complete statistical ensemble assumed in our calculations. Then the observed value of the polarization would correspond to a particular field realization. These values will be fluctuating around the average value derived with the complete ensemble model. The complete ensemble approximation is appropriate if the magnetic field correlation length is well below than the size of the synchrotron emitting layer. The presence of incomplete statistical ensemble of magnetic field fluctuations may result in strong intermittency of the X-ray synchrotron images of SNRs especially when the radiating electrons are in the spectral cut-off regime \citep{Bykov2008,Bykov_Bloemen2009}.

Because of the strong synchrotron losses 
from the magnetic field jump at the subshock,
X-ray emitting electrons are located in a thin layer
near the subshock. 
As a result, the X-ray emission of synchrotron photons of frequency $\nu$ is 
also located in a thin layer of width $d(\nu)$. 
The dependence of this depth, $d$, on  photon frequency, i.e.,  $d\propto1/\sqrt{\nu}$, 
was obtained in a homogeneous magnetic field model in \cite{helder12}, 
where  
different transport regimes of electrons in the emitting layer were discussed. 
We generalised this relation for an inhomogeneous magnetic field decreasing in the downstream direction
and calculated $d(\nu)$ numerically. 
We find for our 
model magnetic field in the X-ray frequency range that $d \propto 1/\nu^I$, 
where $I\approx0.5-1.0$ (depending on $L_0$, the correlation scale of the initial seed magnetic fluctuations). 
 
\begin{figure}[!th]
\includegraphics[viewport=140bp 300bp 590bp 702bp,clip,scale=0.55]{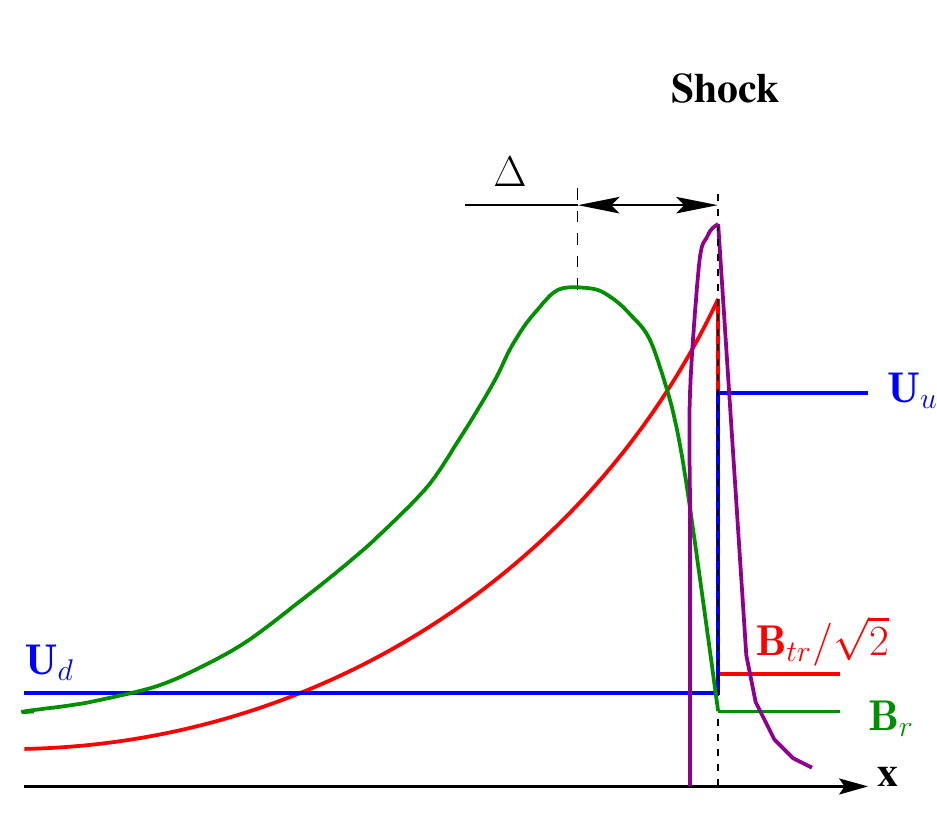}
\includegraphics[viewport=140bp 300bp 590bp 670bp,clip,scale=0.55]{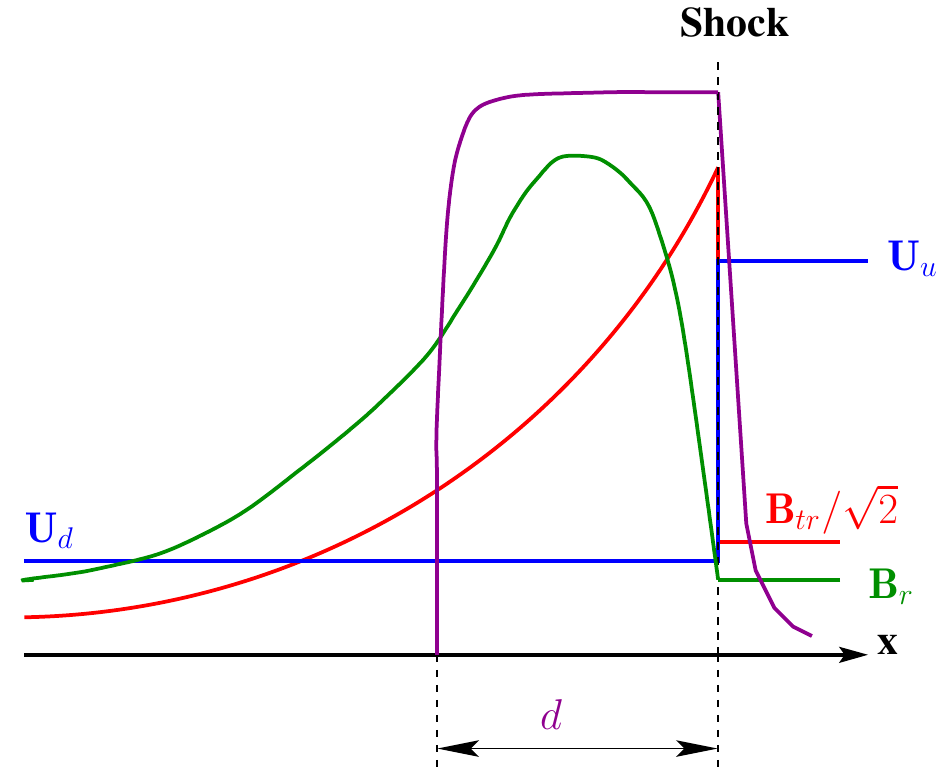}
\caption{In this schematic picture  the upstream region is to the right and the shocked, downstream region is to the left.
 The blue curve shows the plasma flow speed measured in the shock rest frame. 
The RMS amplitude of the radial magnetic field, $B_r$, is shown in green, while the 
transverse magnetic field amplitude $B_\mathrm{tr}$ (divided by $\sqrt{2}$ as explained in the text) is shown in red. 
The peak of the transverse field is at the shock as expected from the compression of the transverse component. 
The turbulent parallel field is amplified by the dynamo mechanism as the plasma flows downstream 
and the peak is shifted from the shock surface to a distance $\Delta$. 
The violet curves are electron 
distributions
peaking within a layer of thickness $d$ downstream. 
X-ray synchrotron emission will peak in regions where both electrons and magnetic field are strong.  
The upper panel illustrates a case where $d \ll \Delta$ and the X-ray polarization  
will be predominantly radial.
The lower panel  illustrates a case where $d > \Delta$ and the X-ray polarization is mainly transverse to the shock normal.
\label{fig:sketch}} 
\end{figure}

\begin{figure}[!th]
\includegraphics[viewport=120bp 350bp 500bp 660bp,clip,scale=0.63]{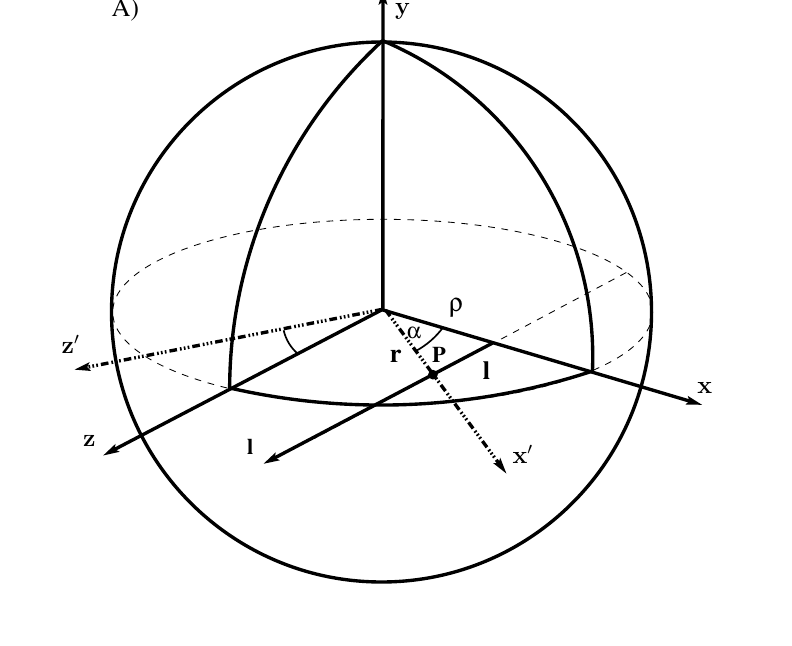}
\caption{Spherical approximation for Tycho's SNR. 
The $z$-axis  and the LOS $l$ are directed to the observer. The $x$-axis is chosen so the LOS
is lying in the x,z plane. The $x^{\prime}$-axis,   that crosses the
LOS at the point $P$, is inclined to the $x$-axis at an angle $\alpha$.
($\rho,l$) are the ($x,z$) coordinates of the point $P$, and $r$ is its distance
to the center of the remnant.  \label{fig:geometry}}
\end{figure}

For a homogeneous  field
($d\propto1/\sqrt{\nu}$) the observable synchrotron spectrum integrated
over the LOS (or over the emission volume) has an index that is greater by $1/2$
than that for the local emission spectrum. 
In this case, the electron
spectral index obtained from the emission spectrum observed from the total emission volume $p_\mathrm{obs}=p+1$.
For the simulated inhomogeneous magnetic field (discussed in Section~\ref{sec:setup}) 
$d \propto 1/\nu^I$ and $p_\mathrm{obs}=p+2I$. 
The position of the peak of the radial  
magnetic field component generated near the shock is shifted by a distance $\Delta$
toward the center of the remnant from the position of the peak
of the tangential field component (see Fig. \ref{fig:sketch}). 

A polarization direction and degree of the X-ray synchrotron emission from 
the SNR ridge is defined by the proportion 
of the magnetic field projections $B_y$ and $B_x$
to the plane transverse to the LOS (see Fig. \ref{fig:geometry} and Appendix A).
If the stochastic magnetic field dominates over the mean field,  
$B_y=B_\mathrm{tr}/\sqrt{2}$. 
For points along the LOS with $\alpha=0$ (Fig. \ref{fig:geometry}),
$B_x=B_r$. 
The components $B_r$ and $B_\mathrm{tr}/\sqrt{2}$ are plotted in Figs.~\ref{fig:sketch} and~\ref{fig:an_approxim}.

If the depth of the X-ray  emitting layer $d\lesssim\Delta$ (bottom panel of Fig.~\ref{fig:sketch}), the total magnetic
field can be considered almost constant over a length $d$ and the relation $p_\mathrm{obs}=p+1$ approximately
holds. 
If $d\ll\Delta$ (top panel of Fig.~\ref{fig:sketch}) the tangential magnetic field dominates and radially
polarized  emission is expected. 
As the width of the electron distribution $d$ increases toward the remnant center, 
the radial magnetic field contribution increases so the polarisation degree decreases at first up to some width $d\sim\Delta$. 
For wider electron distributions $d\gsim\Delta$, the
radial magnetic field dominates and the polarization increases in the tangential direction.
This scenario works if the radial component of turbulent magnetic field 
amplification in the downstream is sufficient. 

The value of the photon spectral index found in \citep{CassamChenHughes07} for the near
ridge regions of Tycho's SNR lies in the interval $2.5-3.2$. 
The corresponding electron indexes $p_\mathrm{obs}\approx4.0-5.4$ and  the allowed local range is $p\approx2.0-4.4$. The local value $p=3.0$ is used in our simulations. 
The uncertainty in $p$ leads to $\sim 15\%$  error in the 
polarization degree calculation which is less than the 
observational error given in \cite{ixpe_Tycho23}.

\section{Setup}
\label{sec:setup}
In this section, we describe the numerical ideal MHD modeling,  
that uses the open code PLUTO \cite{Mignone2007} 
to  
obtain radial profiles of the magnetic field near the shock necessary for calculating synchrotron radiation.
We create dimensionless parameters using the following normalizations:
$\displaystyle\rho_{\ast}=1.67\cdot 10^{-24} \mathrm{g/cm^{3}}$ 
for number density (i.e. 1 proton per $\mathrm{cm}^3$),
$\displaystyle L_{\ast}=5\cdot 10^{16} \mathrm{cm}$ for length, 
and $\displaystyle u_{\ast}=10^{8} \mathrm{cm/s}$ for speed.
The calculations use the ROE solver, parabolic reconstruction, the third-order Runge-Kutta scheme for time steps, and the Eight-Wave Formulation to control the zero divergence of the magnetic field ($\nabla\cdot\mathbf{B}=0$).

The MHD simulation procedure has two distinct stages. In the first
stage, we fix an external cosmic ray current in a box filled with
ambient plasma. This 3D simulation runs for a time corresponding to the
time needed for a plasma parcel to cross the precursor of the shock.
During this time, Bell's turbulence with magnetic field and density
fluctuations is generated. Both the CR current and the scale size of the
shock precursor are consistent with the values obtained in our
non-linear, kinetic Monte Carlo model of shock acceleration in Tycho's SNR \citep{Bykov3inst2014}.
After a shock precursor crossing time, we freeze the turbulence
amplified by Bell's mechanism in the box. The plasma in this first
simulation is then used as a boundary condition for the MHD shock
simulation.

Bell's instability occurs
in the shock precursor and is driven by the 
electric current $\mathbf{j}^{cr}$ of accelerated protons,   
as measured in the upstream plasma rest frame \cite{Bell04,Zirakashvili_2008pol,Bykov2013}.
Large hybrid simulations of magnetic turbulence amplification in DSA \citep{Caprioli_MFA14} demonstrated that Bell's instability grows faster than the resonant instability in shocks with Alfvenic Mach numbers larger than 30. This is the case for the forward shock in Tycho's SNR.

The equation of motion of a perturbed background plasma in 
the same frame can be written as
\begin{equation}\label{eqMotiontot}
\rho\left(\frac{\partial\mathbf{u}}{\partial
 t}+(\mathbf{u}\nabla)\mathbf{u}\right)
 =- \nabla p+  \frac{1}{4\pi}(\nabla\times\mathbf{B})\times\mathbf{B} -
\frac{1}{c}(\mathbf{j}^{cr}\times\mathbf{B}),
\end{equation}
where $\rho$ is the plasma density, $\displaystyle\mathbf{u}$ is the plasma velocity, 
$\displaystyle\mathbf{B}$ is the magnetic field, $p$ is the pressure, and $c$ is the speed of light. 

To initiate Bell's instability, the CR current $\displaystyle\mathbf{j}^{cr}$ 
can be considered constant \cite[see e.g.][]{Bell04,Zirakashvili_2008pol} 
given the small response of the CR current to the short wavelength unstable fluctuations. 
Then, equation Eq.~\ref{eqMotiontot} differs from the 
adiabatic, single-fluid MHD equations solved by the PLUTO code only by an external force term. 
We introduced an external force term driven by the CR current into the momentum and energy balance equations. 
For the initial Bell instability calculation  we use a cubic box containing 64x64x64 cells. 
Our downstream magnetic field MHD simulation (second stage) and polarization calculations use a box with 2560x512x512 cells.

The coordinates for the setup box, normalized to $L_{\ast}$, range from $-1$ to 1 along each side of the box. 
Periodic boundary conditions are set at all box boundaries.
The code is initiated at $t=0$ with the velocity $\displaystyle\mathbf{u}=0$
and $\displaystyle\rho/\rho_{\ast}=0.3$. 
The same  ambient density was used by \citep{2014ApJ...783...33S}
in a non-equilibrium model of broadband emission from Tycho's SNR.  
Our simulation is done in the upstream plasma rest frame. 
Note that there is no shock in the calculation of Bell's instability. 
The CR proton
electric current is directed opposite to the $x$-axis such that $j^{cr}_{x}/j_{\ast}=-4$, $j^{cr}_{y}/j_{\ast}=0$, and $j^{cr}_{z}/j_{\ast}=0$, where the normalization electric current is  
\begin{equation}\label{jast_norm}
j_{\ast}=\frac{c\sqrt{\rho_{\ast}}u_{\ast}}{L_{\ast}\sqrt{4\pi}}.
\end{equation}

The initial magnetic field consists of a constant component directed along 
the $x$-axis equal to $\displaystyle B=3 \mu G$, and a seed turbulent component represented by the sum of modes with a RMS value of $\displaystyle\delta B_\mathrm{rms}=0.1 B$, where 
\begin{equation}\label{Brms}
\delta \mathbf{B}=\sum_{n} a_{n}\mathbf{e}_{n}\cos(\mathbf{k}_{n}\mathbf{r}+\Psi_{n}) .
\end{equation}
Here  $\displaystyle\mathbf{k}_{n}$ is a mode wavenumber, 
$\displaystyle a_{n}\sim k^{-11/6}$ is a mode amplitude (the wavenumber dependence corresponds to  Kolmogorov turbulence), 
$\displaystyle\mathbf{e}_{n}$ is a unit mode polarization vector, and
$\displaystyle\Psi_{n}$ is a mode phase (random). 
The components of the mode wavenumbers are 
\begin{equation}
k_{nx}= \frac{2\pi i}{L}, \; i=0,1,2,..9 \; ,
\end{equation}
\begin{equation}
k_{ny}= \frac{2\pi j}{L}, \; i=0,1,2,..9 \; ,
\end{equation}
\begin{equation}
k_{nz}= \frac{2\pi k}{L}, \; i=0,1,2,..9 \; ,
\end{equation}
where $L$ is the length of the box edge  (in these simulations $\displaystyle L/L_{\ast}=2$), and the sign of the wave vector component is chosen randomly. 
The vector $\mathbf{e}_{n}$ is chosen randomly 
in the plane perpendicular to $\displaystyle\mathbf{k}_{n}$.

In order to setup the amplified magnetic turbulence in the simulation box, 
the simulation is started with initial seed fluctuations and run long enough 
for an upstream plasma parcel to cross the shock precursor. 
In physical units for Tycho's SNR, this time is about 20 yrs, i.e., the time for plasma to flow with the shock speed the
distance of $\sim$ 0.1 pc between the free escape boundary and the viscous shock
\citep[see e.g.][]{Bykov3inst2014}.
During this time, the CR current will amplify the background turbulence via Bell's instability.

\begin{figure}
\includegraphics[scale=0.4]{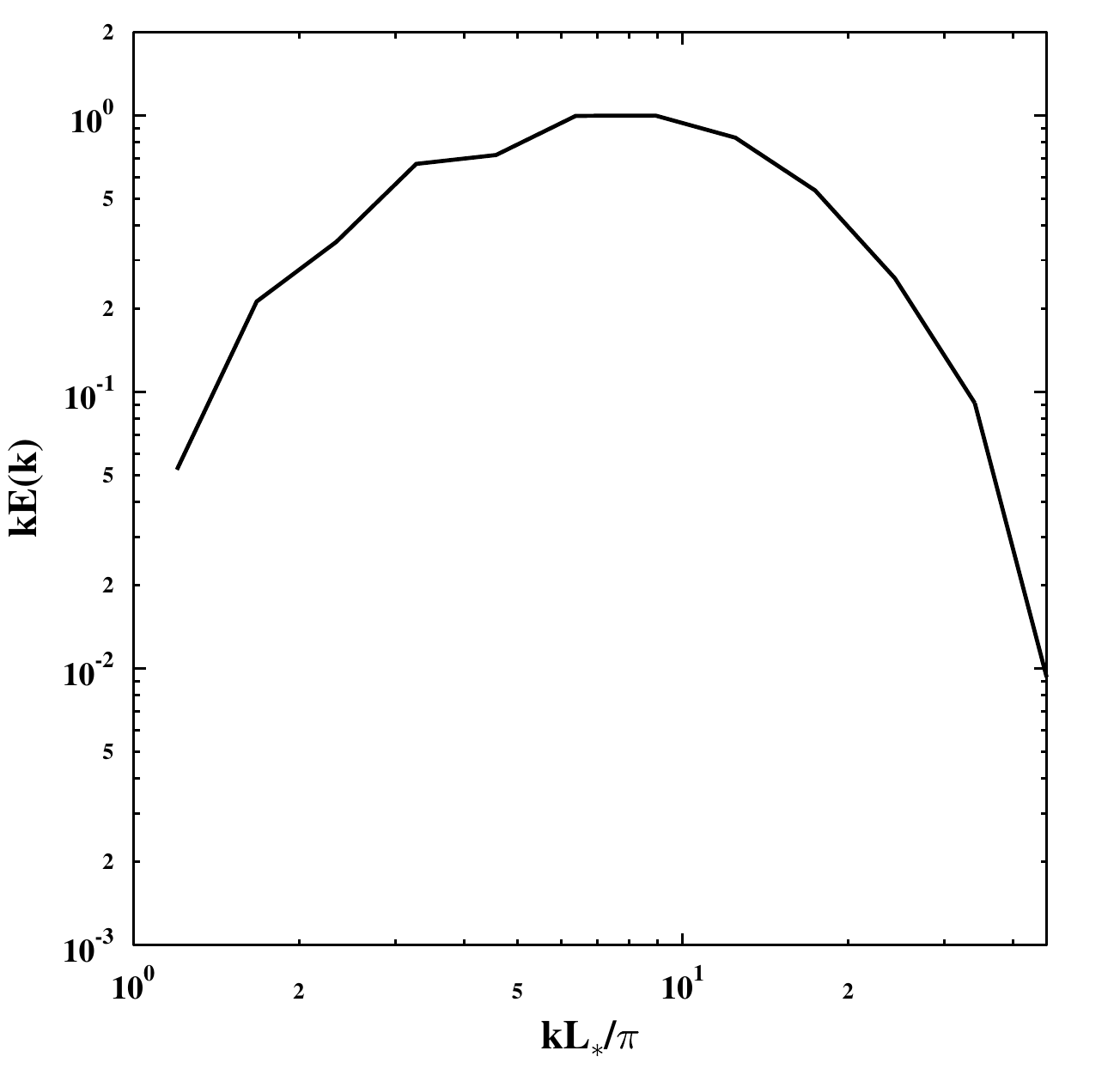}
\caption{Magnetic field energy spectrum normalized to the maximum of the curve 
at the end of the setup simulation for magnetic field amplification by the CR driven instability in the shock precursor. As it is seen the energy containing scale of the turbulence is $\sim L_{\ast}/3$ in this case.} 
\label{energy_sp_Bell}
\end{figure}

At the end of the 
simulation the RMS magnetic field  was $B_\mathrm{rms}\approx 94\, \mu$G. 
The ratio of the RMS turbulent density to the average density was $\approx 0.39$.
In Fig.~\ref{energy_sp_Bell} we show the magnetic field energy spectrum normalized to the maximum of the curve
at the end of the CR driven turbulence simulation.
It can be seen from Fig. \ref{energy_sp_Bell} 
that the turbulence characteristic scale $\displaystyle L_{0}$
is  $\displaystyle kL_{\ast}/\pi\approx 6$, that is $\displaystyle L_{0}=2\pi/k\approx L_{\ast}/3$,
where $L_0$ is of the order of the gyroradius of a CR proton of maximal energy (derived in the amplified field).

The second stage of the calculation is performed 
using the results of the first stage simulation as a boundary condition on the left side of the simulation box. This simulation is done in the shock rest frame. 
The main simulation box has dimensions of 2560x512x512 cells, the plasma flows along the positive $x$-axis, 
and all coordinates are normalized to $\displaystyle L_{\ast}$. The $x$-axis ranges from 0 to 10 over 2560 cells
and the 
perpendicular $y$- and $z$-axes range from \hbox{$-1$ to 1} (as in the previous setup calculation).
Periodic boundary conditions are set for the sides of the box that are parallel to the $x$-axis.
An outflow boundary condition is set on the downstream side of the box at $x/L_{\ast}=10$. 
 A 4,000 \kms~ shock is initiated in the
simulation box. Plasma flows from left to right across the left boundary x=0
at a speed of $\displaystyle 4\cdot 10^{8}$ cm/s, 
 and the plasma crossing  the left boundary is the turbulent plasma from the first
simulation stage (similar to what was 
done in the paper \cite{Zirakashvili_2008pol}).

 Since
the time for  plasma to convect to the downstream wall of the second stage simulation box is
greater than the  size of the simulation domain of the first stage divided by the upstream speed, 
we use a
cycling procedure where we repeat the injection of the Bell turbulent plasma as often as
necessary.
The transformation of the first stage simulation data to the
shock rest frame is taken into account. The interpolation of the 
data obtained at the first stage into the cells
of the left boundary of the main box is performed using methods built into PLUTO. The accelerated proton
current is set to zero so Bell’s instability is inactive. The box  extension
in the x-direction is determined by the grid resolution and calculation time.

The shock initiation was done as follows.
At the initial moment of time, at the point $\displaystyle x/L_{\ast} = 2$, a shock is set: 
for $\displaystyle x/L_{\ast} < 2$, the flow velocity equal to $\displaystyle 4\cdot 10^{8} $cm/s is directed along the $x$-axis, 
the density is equal to the average value from the setup simulation, the magnetic field is directed along the $x$-axis and is 
equal to $\displaystyle 3\mu $G, and the pressure is equal to the average pressure in the box of the first stage calculation. 
With these values, the shock has a Mach number $M\approx 4$.
At $\displaystyle x/L_{\ast}\geq 2$ all values are determined  from the Rankine-Hugoniot conditions, 
so the shock remains almost at rest in the box frame.

\begin{figure*}
\includegraphics[scale=0.6]{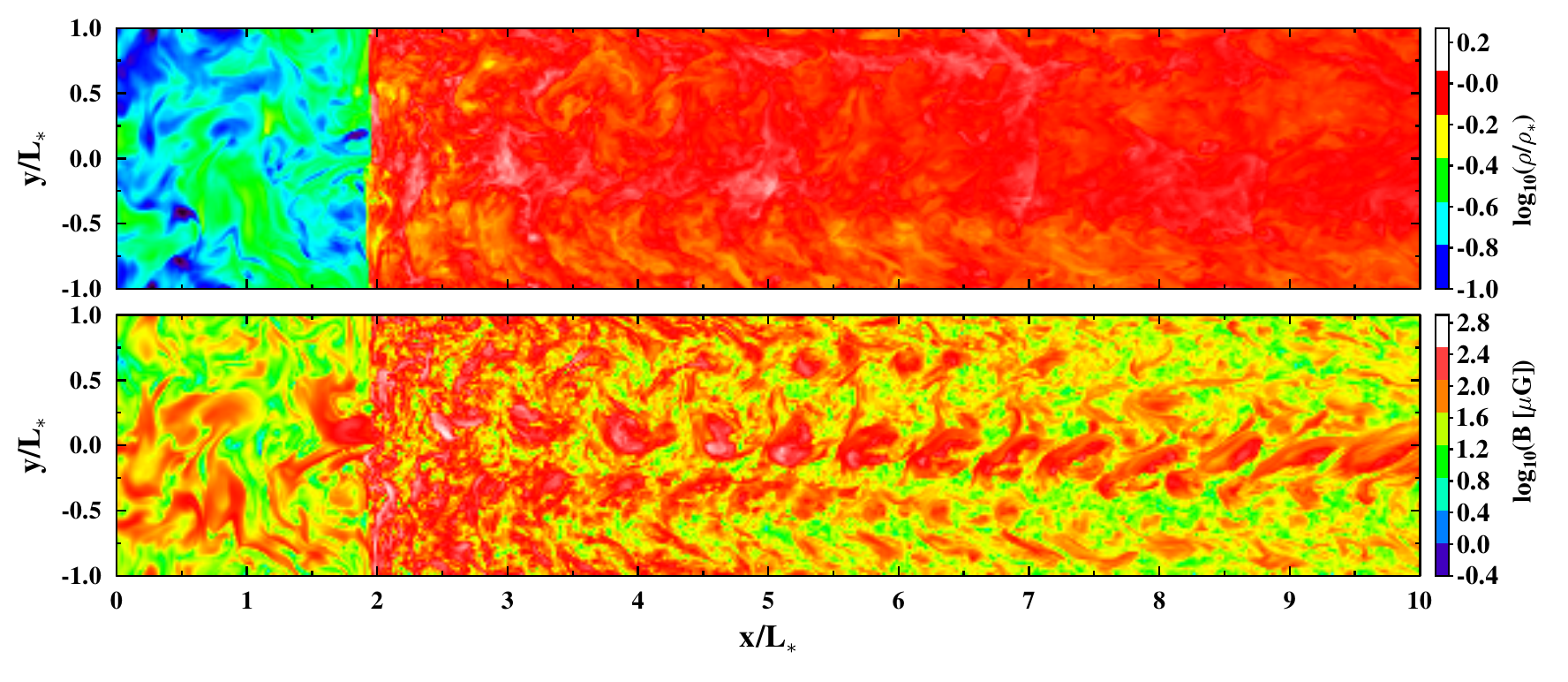}
\caption{Spatial slice of the plasma density (top panel) and the magnetic field modulus (bottom panel) at the end of the simulation for $z=0$.}
\label{Rho_B_2D}
\end{figure*}

\begin{figure}[!th]

\includegraphics[bb= 30 10 570 570,clip,scale=0.47]{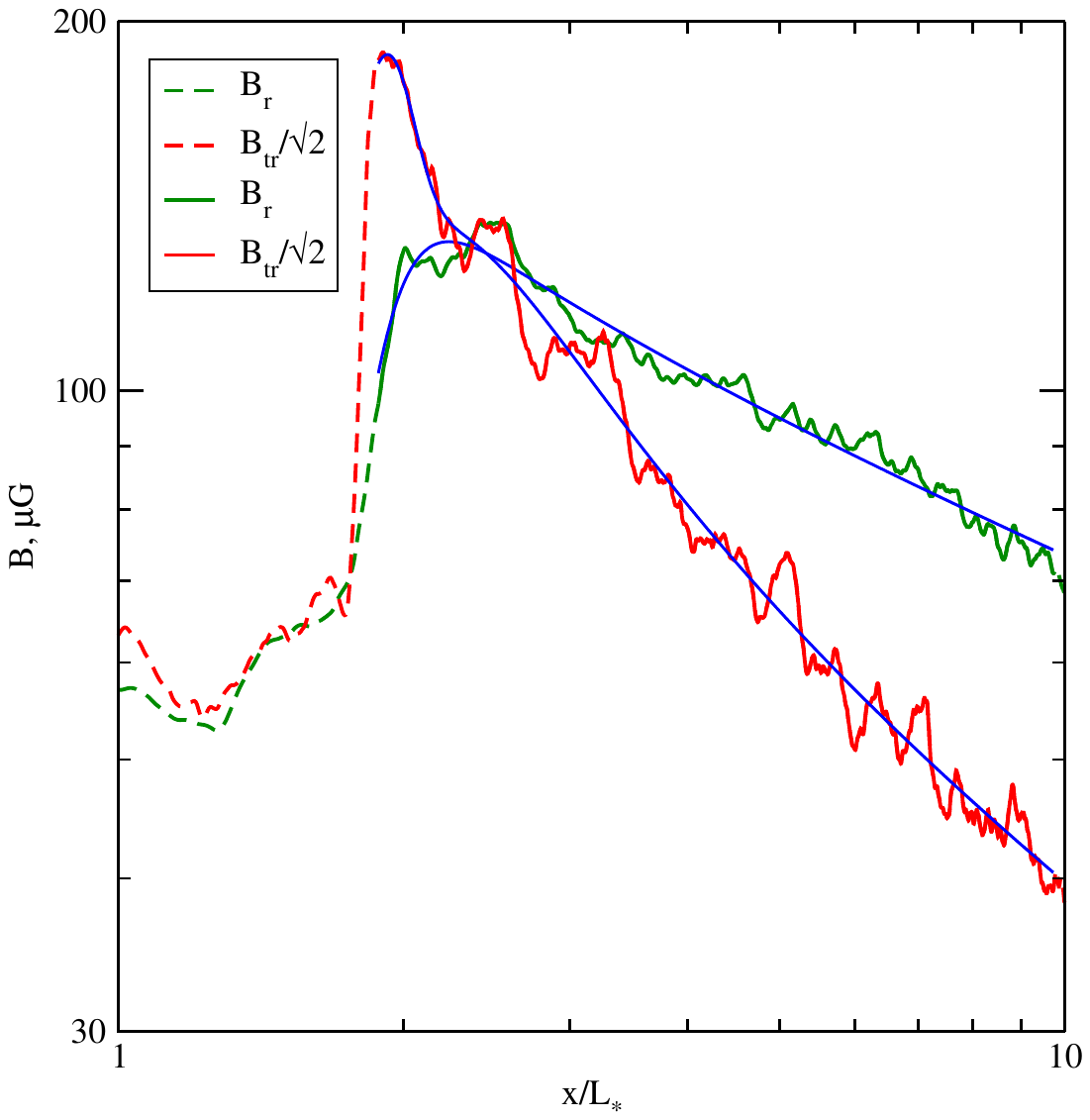}
\caption{
This figure shows the magnetic field projections $B_{r}$ (green)
and $B_{tr}$ (red) mean-square averaged over the simulation box slices with fixed 
$x$-coordinates (in our model the $r$ and $x$-coordinates are identical). 
The analytical approximations fitting the simulation data
are  shown in blue. The fitting is only done over the region shown with solid curves.
In Fig.~\ref{fig:energy_dens} which is plotted in a linear scale we provided a zoom to 
follow the evolution of the energy densities of the most important turbulent components in the vicinity of the viscous velocity jump. 
} 
\label{fig:an_approxim}
\end{figure}

Fig.~\ref{Rho_B_2D} illustrates the spatial distribution of the plasma density and the magnetic field in the computational domain at the end of the simulation.
The RMS magnetic field simulation results
are shown in Fig.~\ref{fig:an_approxim}. 
This mean-square averaging was done over the ($y,z$) plane for each $x$ coordinate. The red curves are the transverse components $B_\mathrm{tr}$,
the green curves are the longitudinal components $B_{r}$, 
and the blue curves are fitted analytic approximations.
As expected, the transverse components show a sharp increase from the density compression 
as the plasma crosses the subshock at $x \approx 2 L_{\ast}$. 
As the plasma continues to flow downstream $B_\mathrm{tr}$ 
decreases.

In contrast, the longitudinal component $B_r$ shows a modest increase immediately at the shock but is 
enhanced by the small-scale dynamo mechanism as the plasma flows downstream. 
At some distance downstream,  
$B_{r}$ becomes, 
and remains, substantially greater than $B_\mathrm{tr}$.

\begin{figure}[!th]
\includegraphics[bb= 5 12 400 390,clip,scale=0.65]{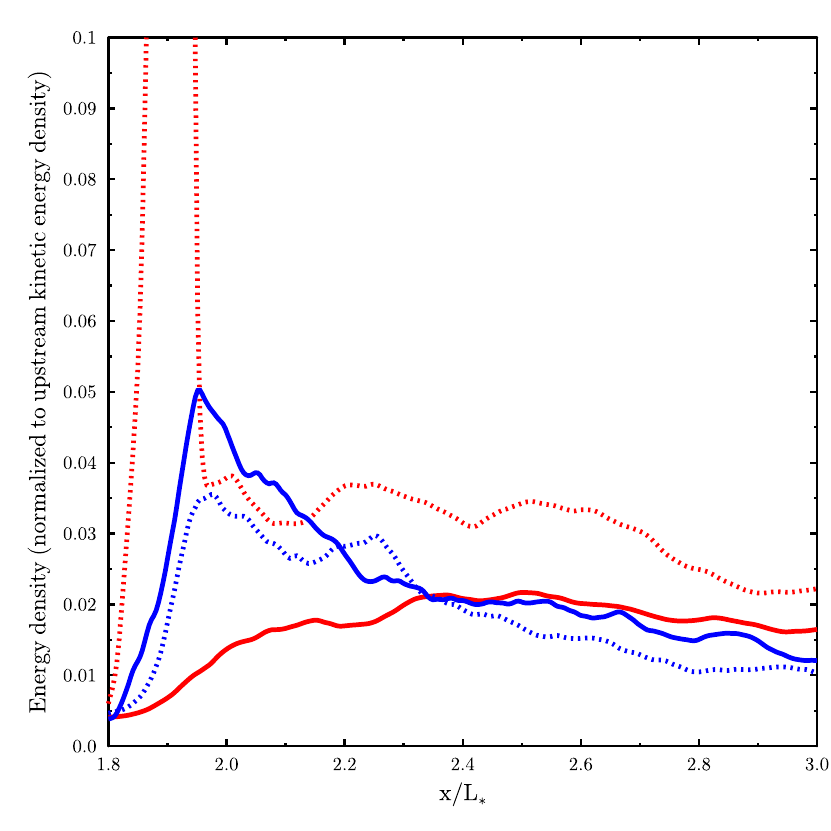}
\caption{Energy density of downstream turbulence normalized to the far upstream kinetic 
energy density. The dotted curves show kinetic energy densities derived from plasma motions while the solid curves show magnetic field values. 
The large peak in the radial component of radial velocity fluctuations (red dotted curve) comes from the corrugation of the shock surface since the averaging 
was done over slices of constant $x$ which intersect both upstream and downstream flows with significantly different speeds.
The width of the peak is about the energy containing scale of the upstream turbulence (see Fig.~\ref{energy_sp_Bell}). The strong shock corrugations drive the radial anisotropy of the downstream turbulence. The radial magnetic field component is the solid red curve while the transverse field component is the solid blue curve. 
The growth of the radial magnetic component and its dominance over the transverse component with distance is seen.}
 \label{fig:energy_dens}
\end{figure}

A unique feature of our model, stemming from the 
delayed dynamo effect, is the spatial inhomogeneity in polarization direction. This is shown in Fig.~\ref{fig:energy_dens}.
The longitudinal magnetic field component is weaker than the transverse component immediately behind the shock  but experiences growth from the dynamo effect as the 
plasma convects away from the shock.  At some point downstream, depending on the SNR parameters, 
the longitudinal component may dominate in the 
main part of the emitting volume. Our model connects the SNR parameters (e.g. shock speed, 
ambient density) to polarization direction, as well as giving a rationale for how anisotropic turbulence can co-exist with efficient DSA.

\section{Simulations vs observations}

We compare our simulation results for Tycho's SNR 
to {\sl Chandra's} X-ray radial intensity profiles obtained for the 4-6 keV energy range \cite{CassamChenHughes07},
and {\sl IXPE's} X-ray polarization maps obtained for the 3-6 keV energy range \cite{ixpe_Tycho23}. 
The \chandra data has near arcsec angular resolution for the 
4-6 keV energy band and is free of emission lines. {\sl IXPE} has much lower angular resolution.
We convolved  our simulation results with the 
{\it Chandra} and {\it IXPE} point-spread-functions (PSFs), 
using the model of \citep{Fabiani_2014ApJS..212...25F} for {\sl IXPE}.
For our comparison, 
we assume a photon energy of 3~keV and the synchrotron fraction in the total emission equal
to be 0.6 for Tycho SNR (see \cite{2020ApJ...899..142B}).  The polarized photon fraction for other young SNRs remnants may be either larger, if the shock is propagating in rarefied plasma (e.g. in a progenitor star wind), or smaller in the case of SNR shells with bright unpolarized thermal emission. Therefore, the polarization degree estimation for these SNRs would differ from the case of Tycho's SNR discussed in the paper.

The \chandra PSF was simulated
using the psfFrac function in Python's psf module 
(http://cxc.cfa.harvard.edu/ciao/ahelp/psf.html) 
for 4 keV photon energy and an offset of 4 arcmin 
(i.e., the Tycho SNR ridge offset
in the \chandra observations used by \citep{CassamChenHughes07}).
For {\sl Chandra}, 80\% of the point source emission at 4 keV
is concentrated in a circle of  $\sim 0.7$ arcsec radius
for the on-axis direction, and in a circle of $\sim 2.3$
arcsec radius for the 4 arcmin off-axis direction. Magnetic field samples
used in the simulation are discussed in Section~\ref{sec:setup}.

\begin{figure}[!th]

\includegraphics[bb= 20 0 560 650,clip,scale=0.46]{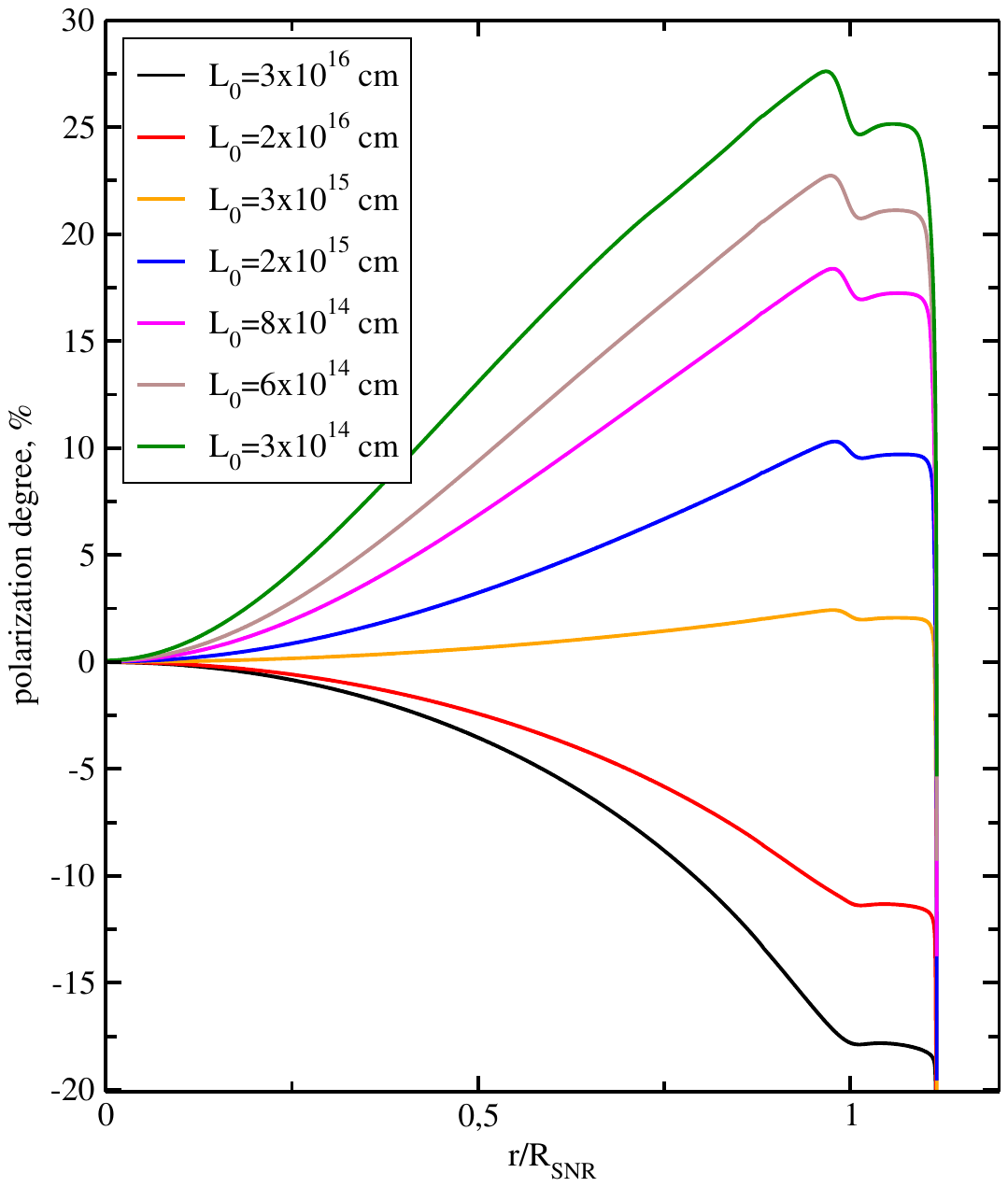}

\caption{X-ray polarization percentages as a function of position downstream 
from the shock for different model magnetic field profiles.
The polarization for different values of $L_0$ is obtained by convoluting the \IXPE PSF for 3 keV photons
for different values of $L_{0}$. The polarized synchrotron photon fraction was assumed to be 0.6, as was estimated earlier for Tycho's SNR (see the text).
Positive
and negative values indicate transverse and radial directions of polarization respectively.
}
\label{fig:XIPE_profile} 

\end{figure}

If the depth of the X-ray emitting layer $d\ll\Delta$, the magnetic
field in the emitting layer is tangent to the shock front and the synchrotron emission
is polarized radially. In the opposite case $d\gtrsim\Delta$,
the magnetic field is radial in most of the emission layer
and the polarization of the total synchrotron radiation is tangential.
In the intermediate case the polarization degree of the total emission
should be low because of the near equal intensity of radial and transverse emission.
This
is illustrated in the lower panel of Fig.~\ref{fig:XIPE_profile} with the \hbox{$L_0 = 3\cdot 10^{15}$~cm} example
($\Delta \propto L_{0}$).

It should be mentioned that the  \ixpe $\sim30$ arcsec PSF radius is greater
than the angular value of $\Delta$ (Fig.~\ref{fig:sketch}) so \ixpe measures the intensity
and polarization from the total emission layer. \citep{ixpe_Tycho23}
found that there is a rather high tangential X-ray polarization
emission from the ridge of Tycho. The polarization degree is  $\sim 20$\%
from the northwest ridge region and $\sim 10$\% overall
(while estimated errors are rather high). 
We conclude from this that $d\gtrsim\Delta$ for Tycho. 
Our simulation results show
that if the polarization is transverse and high ($\ge10\%$),
the value of $L_{0}$ is limited from the upper side, 
i.e  $L_{0}\lesssim 2\cdot 10^{15}\,\mathrm{cm}$ (Fig. \ref{fig:XIPE_profile}).

\begin{figure}[!th]

\includegraphics[bb= 20 50 560 570,clip,scale=0.46]{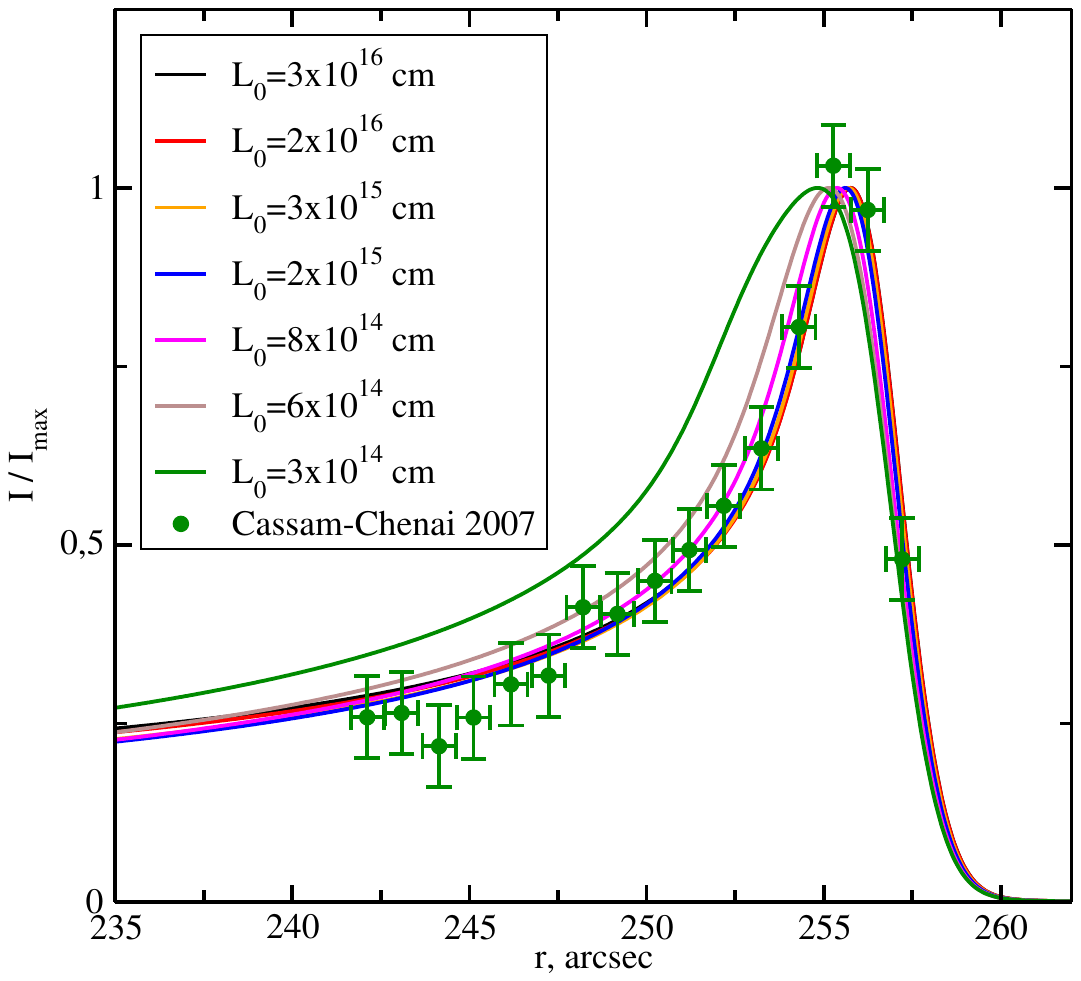}

\caption{Simulated \chandra intensity profiles after convolution with the \chandra
PSF for 4 keV photons for different values of $L_{0}$. 
The data points obtained in \citep{CassamChenHughes07} are shown in
green.   \label{fig:profile}}
\end{figure}

Fig.~\ref{fig:profile} shows the X-ray intensity profile after convolution with the \chandra PSF. 
The data points in green are from \citep{CassamChenHughes07}.
Our results are sensitive
to the value of $d$.
If $d\gg\Delta$, the X-ray \chandra intensity profile
near the ridge is determined by  the 
asymptotically falling downstream magnetic field (see Fig. \ref{fig:an_approxim} and Appendix~\ref{sec:fitting}) and is too broad for a good fit to the data.
However, a high transverse  \IXPE polarization in our model demands that $d\gtrsim\Delta$. This case is shown in the bottom panel of the Fig. \ref{fig:sketch}. 
With decreasing $L_0$ ($L_*$) the depth of the layer with high magnetic field ($\sim\Delta$) decreases as does the gross synchrotron losses in the shock vicinity. Because of the lower loss rate, the width $d$ of the layer of X-ray emitting electrons increases.
If $d$ is less than or equal to the \chandra PSF radius,  the profile
width is defined by the latter and is almost independent of $d$.
This is the reason why in Fig. \ref{fig:profile} there is almost no dependence on
$L_{0}$ if the turbulent scale $L_{0}  \geq 8\cdot 10^{14}$ cm.

For a Tycho SNR distance of 2.5
kpc, 1~arcsec is $\sim4\cdot10^{16}$ cm. The \chandra PSF radius at
the ridge is $\sim2.5$ arcsec so an upper value for $d$  
below which the intensity profile only weakly depends on $d$ is $\sim 10^{17}$ cm.
For the X-ray intensity profiles shown in Fig.~\ref{fig:profile},
only the $L_{0}=3\cdot10^{14}$ cm case has $d\approx 1.6\cdot10^{17}$ cm that 
exceeds the \chandra PSF radius.
The $L_{0}=6\cdot10^{14}$ cm case has $d$ that is approximately equal to it and for all the other plotted curves  $d$ is lower. 
Fig. \ref{fig:profile} shows
that the data points of \citep{CassamChenHughes07}
are well fitted if $d\lesssim \mathrm{PSF}_\mathrm{Chandra}$
and the fit quality worsens if this condition breaks. 
This allows an estimate of the width of the
emission layer to be $d\lesssim \mathrm{PSF}_\mathrm{Chandra}$. 
If besides a good fit of the intensity profile 
we need a high transverse polarisation, then $\Delta\lesssim d$ holds so $\Delta$ is also less than the \chandra PSF radius. 

While the shock propagates in the turbulent medium its front should be
perturbed (see Fig.~\ref{fig:energy_dens}). This effect should increase the observable ridge width after
projection over the LOS and averaging. Such averaging is
used in the Chandra  data processing when counts are gathered from a rather
broad sector of the SNR image to obtain 1-dimension radial profile \citep{CassamChenHughes07}. 
Thus the condition $d\lesssim PSF_\mathrm{Chandra}$
also limits the front ripple amplitude.

\section{Discussion}
Radio and X-ray observations of young SNRs have revealed in a few cases polarized synchrotron radiation produced by predominantly radial magnetic fields. Magnetic turbulence with predominantly radial anisotropy can be produced under certain conditions in the downstream of a strong shock propagating in a turbulent medium (see e.g. \citep{Zirakashvili_2008pol,inoue13,lazarian_MFA22}).
We have described here a physical mechanism for producing  strong turbulence with predominantly radial anisotropy downstream from a SNR shock undergoing efficient DSA and producing high energy CRs. 
We show how strong fluctuations with predominantly radial direction  
are amplified by the anisotropic turbulent plasma motions.
Strong magnetic fluctuations with predominantly radial anisotropy in the downstream of a strong shock are amplified by the small-scale turbulent dynamo mechanism. 

The anisotropic plasma velocity fluctuations behind the shock are produced by upstream density fluctuations flowing into the subshock. These density fluctuations are generated in the shock precursor by the non-linear phase of Bell's instability driven by the electric current of the highest energy CR protons as they escape the accelerator. In our results, density fluctuations of amplitude $\delta \rho/\rho \sim$ 0.4 of scale $L_0 \leq 10^{16}$ cm were produced. As they flow through the subshock, the density fluctuations produce a rippled structure that generates anisotropic vortex turbulence. The anisotropic vortexes behind the rippled shock are due to strong density fluctuations hitting the subshock surface.  
The RMS magnitude of the parallel field component amplified by the small-scale dynamo mechanism reaches a peak at a distance $\sim L_0$ behind the shock. The turbulence then decays, as shown in Fig.~\ref{fig:an_approxim}.  

In principle, the density fluctuations needed to produce the 
parallel 
anisotropy of magnetic turbulence  could be interstellar or circumstellar turbulence \citep{inoue13,lazarian_MFA22}.
However, the amplitude of density fluctuations of scales $\sim 10^{16}$ cm for normal interstellar turbulence associated with SNRs 
is expected to be rather small with $\delta \rho/\rho \sim$ 0.03.
The degree of X-ray transverse polarization in the western part of Tycho's SNR, as measured by \IXPE \citep{ixpe_Tycho23}, is just above 20\%. 
As shown in Fig.~\ref{fig:XIPE_profile}, our simulation results can reproduce this high polarization level if the characteristic 
turbulent scale $L_{0} < 10^{15}$ cm.
Our model assumes strong  magnetic field amplification (well above that due to plasma compression) in the shock precursor in order to match the X-ray profile measured by the \chandra observatory with arcsecond resolution \citep[see e.g.][]{CassamChenHughes07,helder12}.
Cosmic ray driven instabilities \citep[see e.g.][]{SchureEtal2012} can provide the high amplification of seed magnetic fluctuations in the shock precursor as the CR electric current, $J_{CR}$, from shock accelerated protons drives 
Bell's fast non-resonant instability  \citep{Bell04}. 
The wavenumber of the magnetic fluctuations with the fastest growth rate in Bell's linear theory is $\sim k_0/2$, where the characteristic wavenumber of the instability is $k_0 = 4\pi J_{CR}/c B$ \citep{Bell04}. Then, the wavelength of  the fastest growing fluctuation is $l_0 = c B/J_{CR}$. 

In the non-linear Monte Carlo model of CR acceleration by a strong shock, the accelerated CR protons of maximal energy leave the accelerator at the upstream free escape boundary where the magnetic field magnitude, in the case of Type Ia SNRs, likely corresponds to the  interstellar value of a few micro Gauss. 
Assuming that the characteristic length scale $L_0 \sim l_0$ at the free escape boundary, we can estimate the electric current  of maximal energy CRs leaving the accelerator needed to match the polarization observations. 
This current is $J_{CR} \sim 10^{-10}$ CGSE units and is in good agreement 
with the free escape boundary current obtained in the Monte Carlo modeling of DSA \citep{Bykov3inst2014} (see their Fig. 2),
where the model shock parameters closely resembled those of Tycho's SNR.
The maximal energies of the accelerated protons in the model were well above 10 TeV.

The main result of our modeling of Tycho's SNR is that, by combining the 
superb angular resolution of \chandra data with the X-ray polarization measured by {\sl IXPE}, we are able to probe deep into the CR driven magnetic turbulence on scales smaller than the resolution of either telescope.
For example, we show  
that a transverse  polarization of 20\%, 
as measured by \IXPE in the western part of Tycho's SNR,
indicates turbulent scales of $L_0 \lsim 8 \cdot 10^{14}$ cm is needed to fit the data.
The intensity profiles measured with {\sl Chandra}'s resolution are consistent with our model turbulent lengths 
$L_0 \gsim 4 \cdot 10^{14}$ cm. 
Despite that the derived length $L_0$ is less than what can  
be directly resolved by {\sl Chandra}, the polarization data combined with the model allows a look at scales smaller than the instrument resolution. 

The physical reason for this is the overlapping of the narrow layer of 
very high energy electrons accelerated by DSA (thickness $d$) 
with the layer of thickness $\Delta$ filled with magnetic fluctuations of predominantly radial direction, as shown in Fig. \ref{fig:sketch}.
The thickness $d$ decreases with the magnitude of the amplified magnetic field. 
The thickness $\Delta$ depends on the density fluctuations $\delta \rho/\rho$ just upstream from the shock surface.

It should be noted that while here we concentrated on Tycho's SNR, our model should apply equally well to another young SNR SN~1006.
\IXPE detected a high average X-ray polarization of
$\sim$ 20\%  with a nearly parallel magnetic field direction in the north-eastern part of SN 1006  \citep{ixpe_SN1006}.
The outer shock of SN~1006 has an estimated shock velocity $\sim 5,000$ \kms\ but the ambient density is a few times below what we assumed for Tycho. Despite this difference, the model discussed above can explain the X-ray observations of SN 1006 as well.

The maximum amplitude of density fluctuations in the shock precursor
is governed by the level of magnetic turbulence produced by
Bell's instability. For shock velocities $u_s \lsim $ 5,000 \kms, this is $\propto \rho_0 u_s^3$
(see \citep{Bell04} and Fig.11 in \citep{Bykov3inst2014}).
Our simulations indicate that $\Delta$ is shortened with the growth of $\delta \rho/\rho$ so the fastest shocks with the strongest Bell instability would have a short distance 
$\Delta$. A short  $\Delta$ means  
the radial magnetic turbulence would be concentrated near the subshock and dominate over $B_{tr}$ there.
Since the peak amplitude of the radial magnetic turbulence approaches the amplitude of the velocity  of radial turbulence, which dominates the energy density in the downstream region (see Fig.~\ref{fig:energy_dens}), the overall polarization in fast shocks will be transverse.
On the other hand, slower shocks in less dense regions, e.g., in some regions of RX J1713.7-3946, can be expected to show dominantly
transverse magnetic fields in X-ray polarization.

The degree of X-ray polarization measured by \ixpe from the shells of Cas A, Tycho and SN 1006 varies from a few percent in Cas A to $\geq$ 20\% in SN 1006 and Tycho. In all three SNRs, the magnetic field direction is predominantly radial. The ratio  of the polarized synchrotron emission to unpolarized emission fluxes from the hot thermal gas affects the measured polarization degree. In fast forward shocks of young SNRs propagating in rarefied plasma the contribution of unpolarized  thermal X-ray emission may be small and the synchrotron fraction may exceed 0.6, which we estimated for Tycho's SNR. 
Then the degree of polarization can exceed $\sim$25\% which we derived in Fig.~ \ref{fig:XIPE_profile} for Tycho's SNR. 
On the other hand, dense ambient matter and slow velocity shock may reduce the X-ray polarization degree.

It should be noted that we limit ourselves here to  
MHD modeling of Bell's instability initiated by a fixed CR current approximation. The CR current in our model mimics the high energy CRs escaping the acceleration region 
from far upstream of the shock.       
Particle in cell \citep{2020ApJ...905....1H} or kinetic treatments 
\citep{Bykov11} can account for the  CR current response and non-linear saturation effects within the accelerator.   
Presently, the computer resources needed to simulate the multi-scale structure of the extended shock precursor and the evolution of magnetic turbulence in the downstream region, effects important for the problem of interest, are not feasible with the microscopic particle-in-cell technique.

However, magnetic field amplification by CR driven instabilities can be studied with non-linear kinetic Monte Carlo model (see e.g. \citep{Bykov3inst2014}), which use a simplified description of MHD turbulence but account for energy-momentum conservation including the CR current response effects in DSA. The level of magnetic turbulence at the end of the upstream of the fast Tycho's shock dominated by Bell's instability derived in the Monte Carlo simulations is consistent with the turbulent magnetic field magnitude obtained in MHD model described above. This is because the effect of the CR current response is modest in fast shocks with efficient acceleration of very high energy protons escaping the accelerator far upstream.
 
 Despite these limitations, an important prediction of our results is the dependence of the polarization direction on shock speed. 
 MHD \citep{Bell04} and Monte Carlo models \citep{Bykov3inst2014}) have shown that the strength of the CR driven Bell instability and the resulting anisotropic magnetic turbulence in the downstream are higher for faster shocks.  We expect fast SNR shocks will have a high 
 degree of transverse polarization.
Slower shocks in SNRs in 
more rarefied plasma can be expected to show dominantly parallel synchrotron polarization as it is observed in radio observations of a number of SNRs \citep{1987AuJPh..40..771M}.

\section{Conclusions}
We show here that the polarised X-ray radiation  detected by the \ixpe telescope from Tycho's SNR can be  
modeled as the synchrotron radiation of multi-TeV electrons accelerated by diffusive shock acceleration  with magnetic field amplification at the fast forward SNR shock. The model determines the 
degree of polarization and its direction, which corresponds to predominantly radial magnetic fields, consistent with observations.

The non-resonant CR current driven instability \citep{Bell04} amplifies short scale magnetic fluctuations which produce strong density fluctuations in the shock precursor. The interaction of these density fluctuations with the thin collisionless viscous subshock generates intense anisotropic turbulence immediately downstream from the shock. 
As the turbulence decays downstream it amplifies magnetic fluctuations producing a radial dominated anisotropy in fast shocks. 

Strong magnetic turbulence in the shock precursor is required for DSA to 
accelerate particles to well above 10 TeV. Our model shows how high polarization levels can be consistent with strong turbulence.
The thin synchrotron filaments 
detected by \chandra are quantitatively explained in the model.

The combined use of \chandra and \ixpe data allows constraints to be placed on  properties of magnetic turbulence at scales well below the available resolution of modern high energy telescopes. Future imaging of young SNRs 
with the high sensitivity at subarcsecond resolution to be provided with the {\sl Lynx} telescope \citep{Lynx19} 
should lead to a fuller understanding of these CR accelerators including the 
twinkling nature of radiation seen from the shock precursor.

 The mechanism discussed above, while applied here to shocks in SNRs, may be responsible for the production of polarized synchrotron X-ray emission in cosmic ray accelerators with strong MHD shocks produced by fast outflows like powerful winds or jets. The highly amplified anisotropic turbulence in the downstream of fast shocks may provide synchrotron radiation with polarization angle distributions corresponding to the preferential direction of turbulent magnetic field along the shock normal in the sources.

%
\begin{acknowledgments}
We thank the referee for a careful reading of this paper and a report which helped us clarify our results.
S.~O. was supported by the baseline project FFUG-2024-0002 at the
Ioffe Institute. Modeling by A.~B. was supported by the MON grant 23-075-67362-1-0409-000105. Some modeling was performed at the Supercomputing Center of the Peter the Great Saint-Petersburg Polytechnic University and at the Joint Supercomputer Center JSCC RAS. P.~S. acknowledges support from NASA contract NAS8-03060.
\end{acknowledgments}

\appendix
\section{Expressions for synchrotron polarization}
\label{sec:massmatrix_el}
Intensity and polarization of the Tycho synchrotron map can be obtained
after integration of Stokes emission parameters over the LOS
using formulae for synchrotron radiation obtained in \citep{Ginzburg1965}:

\begin{widetext}

\begin{flalign}
I(\nu) & =\frac{\sqrt{3}e^{3}}{mc^{2}}\int dldE\frac{\nu}{\nu_{c}}B_{\perp}\left({\bf r}\right)f_{e}\left(E,{\bf r}\right)\int\limits _{\nu/\nu_{c}}^{\infty}K_{5/3}(\eta)d\eta\nonumber \\
Q(\nu) & =\frac{\sqrt{3}e^{3}}{mc^{2}}\int dldE\frac{\nu}{\nu_{c}}B_{\perp}\left({\bf r}\right)f_{e}\left(E,{\bf r}\right)K_{2/3}\left(\frac{\nu}{\nu_{c}}\right)\cos(2\chi)\\
U(\nu) & =\frac{\sqrt{3}e^{3}}{mc^{2}}\int dldE\frac{\nu}{\nu_{c}}B_{\perp}\left({\bf r}\right)f_{e}\left(E,{\bf r}\right)K_{2/3}\left(\frac{\nu}{\nu_{c}}\right)\sin(2\chi)\nonumber 
\end{flalign}


\noindent where $\nu_{c}=3eB_{\perp}\gamma^{2}/4\pi mc$, $\int dEd\Omega_{E}\cdot f_{e}\left(E,{\bf r}\right)=4\pi\int dE\cdot f_{e}\left(E,{\bf r}\right)=n({\bf r})$.
$I,\,Q,\,{\rm and}\,U(\nu)$ are normalized so the radiation flux
near Earth is given by $dF(\nu)=I(\nu)d\Omega=(dS/r^{2})I(\nu)$ and
so on. The function $f_{e}$ is an isotropic electron distribution function, ${\bf B_{\perp}}$
is a magnetic field projection to a plane transverse to the LOS, and $\chi$ is the angle between the fixed direction
in this plane and the main axis of the polarization ellipse. The parameters
$\nu_{c}$ and $\chi$ are functions of ${\bf r}$.

Assuming a power law approximation for the electron distribution function, the integration over electron energy can be done analytically,
see \citep{Ginzburg1965}:

\begin{flalign} \label{eq:I_power}
\widetilde{I}({\bf r},\nu) & =W_{0} \frac{p+7/3}{p+1} B_{\perp}^{(p+1)/2}\nu^{-(p-1)/2}\nonumber \\
\widetilde{Q}({\bf r},\nu) & =W_{0} \cos(2\chi)
B_{\perp}^{(p+1)/2}\nu^{-(p-1)/2}
\end{flalign}
\[
W_{0}
=\frac{\sqrt{3}e^{3}K_{e}}{4mc^{2}R^{2}}\left(\frac{3e}{2\pi m^{3}c^{5}}\right)^{(p-1)/2}\Gamma\left(\frac{3p-1}{12}\right)\Gamma\left(\frac{3p+7}{12}\right) 
\]
here $I(\nu)=\int\widetilde{I}({\bf r},\nu)R^{2}dRd\Omega$ and so
on, $R$ is a distance from the point at LOS to the observer.
The local averaging of eq. (\ref{eq:I_power}) over the stochastic magnetic
field is done assuming the Gaussian probability distribution function
(PDF):
\begin{flalign} 
\label{eq:PDF-1}
dP & =P\left(B_{x},B_{y}\right)dB_{x}dB_{y}
=\frac{1}{2\pi\sigma_{x}\sigma_{y}}exp\left(-\frac{B_{x}^{2}}{2\sigma_{x}^{2}}-\frac{B_{y}^{2}}{2\sigma_{y}^{2}}\right)dB_{x}dB_{y}=\nonumber \\
 & =\frac{1}{2\pi\sigma_{x}\sigma_{y}}exp\left(-\frac{B^{2}cos^{2}\left(\phi\right)}{2\sigma_{x}^{2}}-\frac{B^{2}sin^{2}\left(\phi\right)}{2\sigma_{y}^{2}}\right)BdBd\phi  
\end{flalign}
The angle $\phi$ is measured counterclockwise from the $Ox$ axis and $\sigma_{x}^{2}=\left\langle B_{x}^{2}\right\rangle $, $\sigma_{y}^{2}=\left\langle B_{y}^{2}\right\rangle $.
We measure angle $\chi$ counterclockwise from the $-Ox$ axis (Fig. \ref{fig:geometry}) so $\chi=\phi+\pi/2$, $\cos\chi=-B_{y}/B_{\perp}$, $\sin\chi=B_{x}/B_{\perp}$ and the local emission at point P:
$\widetilde{Q}({\bf r},t,\nu)\propto\left(B_{y}^{2}-B_{x}^{2}\right)/B_{\perp},\;\widetilde{U}({\bf r},t,\nu)\propto2B_{y}\cdot B_{x}/B_{\perp}$.
After averaging over the magnetic field directions one obtains $\left\langle \widetilde{U}({\bf r},t,\nu)\right\rangle =0$.
This means that $U=0,$ and the value of the polarization degree
is ${\Pi}=\left| Q \right| /I$. This is a consequence of the problem symmetry and the
choice of coordinate axes.
The parameters $\sigma_{x}^{2}=\left\langle B_{x}^{2}\right\rangle $,
$\sigma_{y}^{2}=\left\langle \boldsymbol{B_{tr}^{2}}\right\rangle /2$,
$q=2\sigma_{y}^{2}/\sigma_{x}^{2}$ are functions of the position
at LOS. $\sigma_{y}^{2}=\left\langle \boldsymbol{B_{tr}^{2}}\right\rangle /2$,
$\left\langle B_{r}^{2}\right\rangle $, $\left\langle \boldsymbol{B_{tr}^{2}}\right\rangle $
depend on $r$.  $\sigma_{x}^{2}=\left\langle B_{r}^{2}\right\rangle cos^{2}\left(\alpha\right)+\left\langle \boldsymbol{B_{tr}^{2}}\right\rangle sin^{2}\left(\alpha\right)/2$,
$q=2\left\langle B_{y}^{2}\right\rangle/\left\langle B_{x}^{2}\right\rangle=2q_{0}/\left(2cos^{2}\left(\alpha\right)+q_{0}sin^{2}\left(\alpha\right)\right)$
 also depend on $\alpha$ -- an inclination angle shown on Fig. 1,
$tg\alpha=l/\rho$, $q_{0}=\left\langle \boldsymbol{B_{tr}^{2}}\right\rangle /\left\langle B_{r}^{2}\right\rangle $.

\begin{flalign*}
\left\langle \widetilde{I}({\bf r},\nu)\right\rangle  & =W_{0}\frac{p+7/3}{p+1}\int\limits _{0}^{\infty}\int\limits _{-\pi}^{\pi}\frac{dB_{\perp}d\phi }{2\pi\sigma_{x}\sigma_{y}} B_{\perp}^{(p+3)/2} \nu^{-(p-1)/{2}} exp{\left(-\frac{B_{\perp}^{2}cos^{2}\left(\phi\right)}{2\sigma_{x}^{2}}-\frac{B_{\perp}^{2}sin^{2}\left(\phi\right)}{2\sigma_{y}^{2}}\right)}=\\
 & =\frac{W_{0}}{4\pi\sigma_{x}\sigma_{y}}
 \frac{p+7/3}{p+1}
 { \Gamma\left(\frac{p+5}{4}\right)}
 \left(\frac{\sigma_{y}^{2}+\sigma_{x}^{2}}{4\sigma_{x}^{2}\sigma_{y}^{2}}\right)^{-(p+5)/{4}}
 \nu^{-(p-1)/2}
 \int\limits _{-\pi}^{\pi}d\phi\left(1-\frac{\sigma_{y}^{2}-\sigma_{x}^{2}}{\sigma_{y}^{2}+\sigma_{x}^{2}}cos\left(\phi\right)\right)^{-(p+5)/{4}}
 \end{flalign*}
 
 \begin{flalign*}
\left\langle \widetilde{Q}({\bf r},\nu)\right\rangle  & =W_{0}\int\limits _{0}^{\infty}\int\limits _{-\pi}^{\pi}\frac{dB_{\perp}d\phi 
}{2\pi\sigma_{x}\sigma_{y}}
B_{\perp}^{\frac{p+3}{2}} \nu^{-(p-1)/{2}} cos\left(2\phi+\pi\right)
exp{\left(-\frac{B_{\perp}^{2}cos^{2}\left(\phi\right)}{2\sigma_{x}^{2}}-\frac{B_{\perp}^{2}sin^{2}\left(\phi\right)}{2\sigma_{y}^{2}}\right)}=\\
 & =\frac{W_{0}}{4\pi\sigma_{x}\sigma_{y}}
 \Gamma\left(\frac{p+5}{4}\right)\left(\frac{\sigma_{y}^{2}+\sigma_{x}^{2}}{4\sigma_{x}^{2}\sigma_{y}^{2}}\right)^{-(p+5)/{4}}
 \nu^{-(p-1)/{2}}
 \int\limits _{-\pi}^{\pi}d\phi \, cos\left(\phi\right)
 \left(1-\frac{\sigma_{y}^{2}-\sigma_{x}^{2}}{\sigma_{y}^{2}+\sigma_{x}^{2}}cos\left(\phi\right)\right)^{-(p+5)/{4}}
\end{flalign*}

Taking into account that $q=q(\alpha,r)=2\sigma_{y}^{2}/\sigma_{x}^{2}$ one obtains:
\begin{flalign}
\left\langle \widetilde{I}({\bf r},\nu)\right\rangle  & 
=\frac{W_{0}}{\pi 2 \sqrt{2}}
\frac{p+7/3}{p+1}
{\Gamma\left(\frac{p+5}{4}\right)\sigma_{y}^{(p+1)/{2}}\sqrt{q}\left(\frac{q+2}{8}\right)^{-(p+5)/{4}}}
\nu^{-(p-1)/{2}}
\int\limits _{0}^{\pi}d\phi\left(1-\frac{q-2}{q+2}cos\left(\phi\right)\right)^{-\frac{p+5}{4}}\label{eq:local_I} \nonumber \\
\left\langle \widetilde{Q}({\bf r},\nu)\right\rangle  & =\widetilde{\Pi}\left(p,q\right)\left\langle \widetilde{I}({\bf r},\nu)\right\rangle \\
\widetilde{\Pi}\left(p,q\right) & =\frac{p+1}{p+7/3}\frac{\int\limits _{0}^{\pi}d\phi\left(1-\frac{q-2}{q+2}cos\left(\phi\right)\right)^{-(p+5)/{4}}cos\left(\phi\right)}{\int\limits _{0}^{\pi}d\phi\left(1-\frac{q-2}{q+2}cos\left(\phi\right)\right)^{-(p+5)/{4}}}\nonumber 
\end{flalign}
The polarization degree $\left|\widetilde{\Pi}\left(p,q\right)\right|$
(\ref{eq:local_I}) of the local contribution is a function of only
2 parameters: q and p. 

\section{Fitting procedures}
\label{sec:fitting}
The magnetic field obtained in Section~\ref{sec:setup} remains irregular
even after averaging over slices of the simulation
box with particular $x$-coordinates. The result of such averaging in the downstream region
is shown with red and green curves in Fig.~\ref{fig:an_approxim}, together with a smooth
analytic approximations (blue curves). The following
functions were used for the approximation:

\begin{eqnarray*}
B_{r,\mu G} & = & \frac{A_{0} \cdot \left[1+\left(\frac{x-A_{3}}{A_{1}}\right)^{2}\right]^{A_{2}} }{1+A_6\cdot exp\left(-\frac{x-A_{4}}{A_{5}}\right)}\\
B_{tr,\mu G} /\sqrt2& = & C_{0}\cdot \left[1+\left(\frac{x-C_{3}}{C_{1}}\right)^{2}\right]^{C_{2}} + C_{6}\cdot exp \left[ -\left(\frac{x-C_{4}}{C_{5}}\right)^{2}\right] .
\end{eqnarray*}

The fitting was done in the downstream region with variable coefficients
$A_{i}$ and $C_{i}$. The coefficient values obtained
for $\chi^{2}$ minimization are: $A_0=230.0$, $A_1=0.25$, $A_2=-0.16$, $A_3=0.85$, $A_4=0.0$, $A_5=0.13$, $A_6=2.2\cdot10^5$,
$C_0=137.6$, $C_1=0.75$, $C_2=-0.26$, $C_3=1.95$, $C_4=1.80$, $C_5=0.17$, and $C_6=51.5$.
All spatial dimension variables
are measured in units of $L_{*}$ with $x=0$ being the simulation box boundary. 
The shock position 
at the end of the simulation is located at $x=1.8L_{*}$. 
\end{widetext}

\section*{References}

\bibliographystyle{apsrev}

\end{document}